# Scatterer induced mode splitting in poly(dimethylsiloxane) coated microresonators

Lina He, Sahin Kaya Ozdemir, Jiangang Zhu, and Lan Yang[a]

Department of Electrical and Systems Engineering, Washington University,

St. Louis, Missouri 63130, USA

**Abstract** We investigate scatterer induced mode splitting in a composite microtoroidal resonator ($Q \sim 10^6$) fabricated by coating a silica microtoroid ($Q \sim 10^7$) with a thin poly(dimethylsiloxane) layer. We show that the two split modes in both coated and uncoated silica microtoroids respond in the same way to the changes in the environmental temperature. This provides a self-referencing scheme which is robust to temperature perturbations. Together with the versatile functionalities of polymer materials, mode splitting in polymer and polymer coated microresonators offers an attractive sensing platform that is robust to thermal noise.

---

[a] yang@ese.wustl.edu; URL: http://www.ese.wustl.edu/~yang/

Polymer or polymer coated optical microresonators have attracted increasing interest in recent years due to their advantages (e.g., low cost, diversity, easy fabrication, flexible mechanical properties, rich surface functionalities, and easy incorporation of functional materials) over semiconductor and inorganic materials. Poly(methyl methacrylate) (PMMA), Polystyrene (PS), Poly(dimethylsiloxane) (PDMS) and SU-8 are the most commonly used polymers for fabricating microresonators because of their good optical properties.[1-6] Polymer microresonators supporting whispering gallery modes (WGMs) have been utilized to detect bio-molecules, chemicals,[7] temperature,[4] humidity,[8] ultrasound wave,[9] and mechanical force.[10,11] These resonator-based sensing techniques count on monitoring the frequency shifts of the resonant modes, which result from changes in effective refractive index of the resonant modes or deformation of the polymer structure. In addition, composite microresonators formed by coating polymer on semiconductor or inorganic resonators are proposed to reduce the effect of thermal drift on resonator-based sensors. This is achieved due to the negative thermo-optic coefficients of polymers which compensate for the positive thermo-optic coefficients of inorganic materials.[5] Moreover, thin polymer coating layer enhances extension of the evanescent field into the surrounding medium, thereby increasing the sensing volume of the WGM sensor.[12] Although the polymer layer in a composite resonator helps reduce thermal drift, the coating thickness is not easy to control and the thermal effect can be eliminated only for some specific WGMs. On the other hand, pure polymer resonators may suffer significantly from the thermal noise as a result of their large negative thermo-optic coefficients.

Recently, we proposed mode splitting in a high-$Q$ microresonator as an alternative mechanism to resonance shift for sensing applications, and demonstrated highly-sensitive detection and size measurement of single nanoparticles down to 30 nm in radius using a silica microtoroidal resonator.[13] The resonance frequency splitting is a result of lifted degeneracy of counter-propagating WGMs in a resonator due to backscattering induced coupling between them.

This is reflected as a transition from a single WGM resonance to a doublet in the transmission spectrum. Split resonances introduced by a subwavelength spherical scatterer of radius $R$ can be characterized by the doublet splitting $2g = -\alpha f^2(\mathbf{r})\omega_c / V_c$ and the linewidth difference of the two split modes $2\Gamma_R = \alpha^2 f^2(\mathbf{r})\omega_c^4 / (3\pi v^3 V_c)$, where $f(\mathbf{r})$ represents the normalized WGM distribution at the location $\mathbf{r}$ of the scatterer, $\omega_c$ is the angular resonance frequency, $v$ denotes the speed of light in the surrounding medium, and $V_c$ is the mode volume. Polarizability $\alpha$ of the scatterer is calculated by $\alpha = 4\pi R^3 (n_p^2 - n_m^2) / (n_p^2 + 2n_m^2)$, where $n_p$ and $n_m$ denote the refractive indices of the scatterer and the surrounding medium, respectively. As both $g$ and $\Gamma_R$ can be measured from the mode splitting spectrum, radius $R$ of the scatterer is estimated as

$$R = \left[ \frac{3\lambda_c^3 \cdot (\Gamma_R / g)}{32\pi^3 n_m^3 (n_p^2 - n_m^2)/(n_p^2 + 2n_m^2)} \right]^{1/3} \quad (1)$$

where $\lambda_c$ is the resonance wavelength. Since the two split modes reside in the same resonator and their differences in resonance frequencies and linewidths are used to extract the polarizability of the detected particle, a self-referencing scheme is formed where one of the split modes acts as a reference to the other mode and vice versa.[13] This is suggested as one of the possible reasons for the enhanced sensitivity of particle detection using the mode splitting technique.

In this Letter, we present mode splitting in a polymer-silica composite microresonator, and verify the self-referencing issue in mode splitting based sensing schemes by studying the resonance splitting in PDMS coated and pure silica microtoroidal resonators. Responses of the split modes to temperature variations are studied theoretically and experimentally, demonstrating that temperature changes in the environment have a negligible effect on mode splitting.

Silica microtoroids in our experiments are prepared through photolithography technique followed by $CO_2$ laser reflow.[14] Composite resonators are obtained by coating silica microtoroids with PDMS layers using wetting technique. When a PDMS prepolymer droplet is brought into contact with the periphery of a silica microtoroid, it spreads out along the toroid surface due to the low surface tension of PDMS. The coating thickness can be controlled by the size of the droplet.[5] The silica microtoroids have $Q$ above $10^7$ and diameters around 50 μm. After PDMS coating, $Q$ factors are above $10^6$, mainly limited by the material loss of PDMS. A tunable external-cavity diode laser and a function generator are utilized to excite WGMs in the microresonator. Light is coupled into and out of the resonator via a fiber taper. Transmitted light is detected by a photodetector connected to an oscilloscope for monitoring the transmission spectrum. In this study, we intentionally introduce resonance splitting in a PDMS coated microtoroid by gradually approaching a sharp fiber tip (i.e., external Rayleigh scatterer) towards the periphery of the microtoroid. During this process, we continuously monitor changes in the transmission spectrum. The cone-like shape of the tip allows us to simulate a scattering center with an increasing size interacting with the WGM field. Introducing a scatterer (fiber tip) induces the resonance frequency splitting if previously there is no observable splitting for the WGM of interest (Fig. 1(a)).[15] Due to surface defects introduced by non-ideal coating processes or contaminations induced by dust and other particulates, which lead to inhomogeneities and irregularities in the PDMS layer, mode splitting appears at specific resonance wavelength if $Q$ of the WGM is sufficiently high. Such resonance splitting associated with the device itself is termed as intrinsic mode splitting and is presented in I of Fig. 1(b). Changes of mode splitting spectra with increasing tip size are shown in Fig. 1(b).[15] Approximating the fiber tip to a sphere and using Eq. (1), we estimate the radius of the tip interacting with the WGM in Fig. 1(a) as 139.9 nm (II), 185.4 nm (III) and 227.1 nm (IV), and in Fig. 1(b) as 99.6 nm (II), 197.3 nm (III) and 230.6 nm (IV), respectively. Figure 1(c) depicts a fiber tip approaching the periphery of a microtoroid. It is worth noting that one can introduce scatterers to the composite resonator either by directly depositing them onto the PDMS coating surface or by embedding them into the PDMS layer during the coating process. Field distributions of the resonator WGMs obtained from numerical simulations for both cases are presented in Figs. 1(d) and 1(e), respectively.

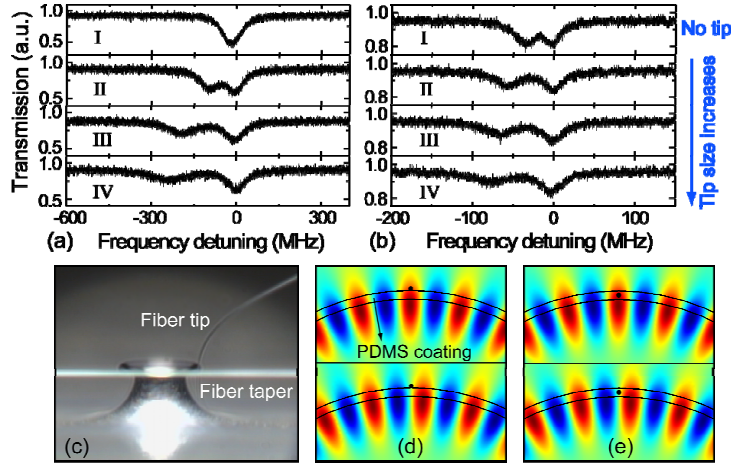

Fig. 1. Mode splitting in a PDMS coated silica microtoroid. Transmission spectra of WGMs at 1437.8 nm (a) and 1428.2 nm (b) in response to a scatterer (fiber tip) with increasing size, indicated by the downward-pointing arrow. Linewidth of the resonant dip in I of (a) is 80.1 MHz corresponding to Q of 2.6 106, while the initial mode splitting in I of (b) is 33.3 MHz obtained through a Lorentz fit. (c) Micrograph of a fiber tip approaching a fiber taper coupled microtoroid. Field distribution of the symmetric mode (upper panel, particle located at the anti-node of the mode) and asymmetric mode (lower panel, particle located at the node of the mode) with a particle deposited on the PDMS surface (d) and doped inside the PDMS layer (e).

To demonstrate the self-reference property of mode splitting, we take the change in environmental temperature as an example of undesirable perturbations and study its effect on the split modes in resonators. As temperature varies, resonances of both split modes shift, as a result of the thermally induced changes in refractive index and size of the resonator. Intuitively, one also expects that the thermal changes modify $2g$ and $2\Gamma_R$ via their dependence on $\alpha$ and $\omega_c$, both of which are influenced by temperature. Numerical simulations for a combined system of a PDMS coated silica microtoroid and a silica scatterer reveal that for a temperature increase of 10 ºC, changes in $2g$ and $2\Gamma_R$ are of kHz order, much smaller than the 10 GHz shift in $\omega_c/2\pi$ (Fig. 2(a)). Note that if the thermal effects on the polarizability of the scatterer are neglected and only the effects on the resonator are considered, the variations in mode splitting are even smaller than the values depicted in Fig. 2(a).

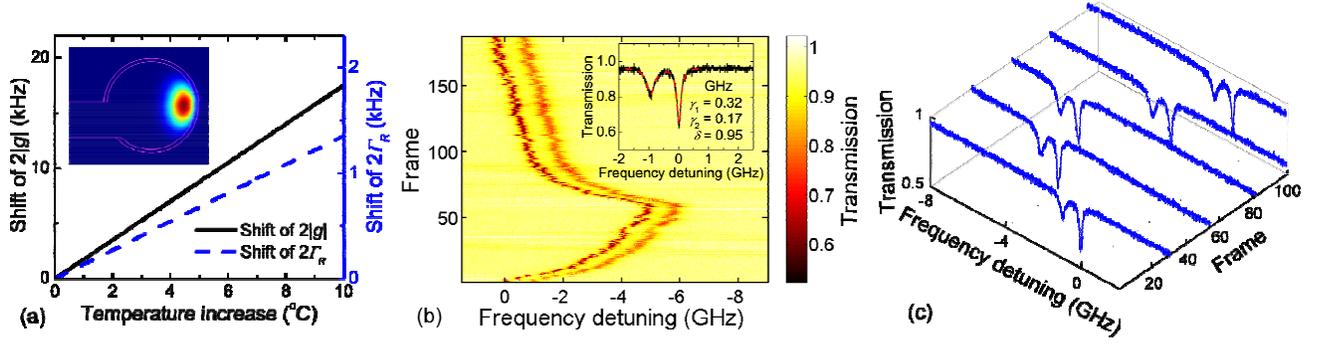

Fig. 2. Results of numerical simulations and experiments. (a) Mode splitting in a PDMS coated silica microtoroid as a function of temperature. The original values for $2|g|$ and $2\Gamma_R$ are 90.4 MHz and 14.5 MHz, respectively. Parameters used in the simulations are: Major (minor) diameter of the silica microtoroid: 45 μm (5 μm); PDMS coating thickness: 0.2 μm; Refractive indices of silica and PDMS: $n_{silica}$ = 1.45, $n_{PDMS}$ = 1.41; $R$ = 130 nm; $\lambda_c$ = 1548 nm; $V_c$ = 358 μm³; $f(r)$ = 0.15; Thermo-optic (-expansion) coefficient of silica: $1.19\times10^{-5}$ °C⁻¹ ($5.5\times10^{-7}$ °C⁻¹); Thermo-optic coefficient of PDMS: $-1.0\times10^{-4}$ °C⁻¹. Inset: cross-sectional mode distribution. (b) Experimentally obtained intensity graph of the transmission spectrum from a PDMS coated microtoroid showing the response of mode splitting spectrum to thermal changes. Inset: the WGM is excited around 1536.5 nm with linewidths $\gamma_1$, $\gamma_2$ and splitting $\delta$ = 950 MHz induced by a fiber tip. (c) Transmission spectra at different frames. Increasing frame number in (b) and (c) corresponds to increasing time.

In the experiments, we used a fiber tip to introduce mode splitting in a PDMS coated microtoroid. Temperature was controlled using a thermoelectric cooler (TEC) placed under the microtoroid chip. Figure 2(b) presents a typical intensity graph of the split resonances, where the TEC was turned on at Frame # 1 to start heating the chip and turned off at Frame #58 to let the chip cool down. Transmission spectra from the oscilloscope are acquired continuously at a speed of 30 frames/min during the process. Mode splitting spectra at various frames are plotted in Fig. 2(c). It is clearly seen that resonance wavelengths of both modes of the doublet experience red (blue) shifts as temperature increases (decreases). During the heating up and cooling down process, no significant alterations are observed in $2g$ or $2\Gamma_R$. Statistical analysis of Fig. 2(b) shows that the amount of splitting is 953 ± 25 MHz, while the linewidth difference of the split modes is 170 ± 26 MHz. Fluctuations in mode splitting can be attributed to the thermally induced changes in refractive index and size of the fiber tip, resonance shift of the resonator, perturbations of the tip position with respect to the mode volume, and variations of the taper-cavity gap. It should be noted that the thickness of the PDMS layer mainly determines the rate of change in resonance frequencies of the doublet modes with temperature (i.e., slope of the curve in Fig. 2(b)), but does not affect the conclusions drawn above for temperature effects on mode splitting.

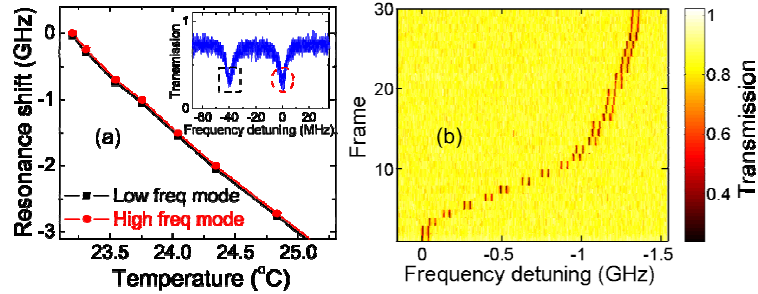

Fig. 3. Experimentally obtained mode splitting spectra for a silica microtoroid. (a) Resonance shift of the split modes as a function of temperature. The red dots and black squares denote the high frequency and low frequency modes in the transmission spectrum, respectively, as marked in the inset. Inset: Mode splitting spectrum (doublet splitting ~ 40 MHz) at 1437.8 nm. (b) Intensity graph of mode splitting spectra as temperature increases.

Compared to polymer-silica composite microresonators, pure silica resonators have superior capability to resolve small changes in mode splitting due to their higher $Q$ factors. Here we study the temperature response of mode splitting in silica microtoroids as a reference to further verify the stability of the split resonances against thermal perturbations. Mode splitting is introduced by depositing a particle on a silica microtoroid (Inset in Fig. 3(a)). In Fig. 3(a), we see that as temperature increases, resonances of both split modes shift to longer wavelength linearly, with the rate of shift (slope of the curve) mainly determined by the thermo-optic coefficient of silica. Figure 3(b) presents the evolution of mode splitting with temperature while the TEC is turned on at Frame #1. The transmission spectra are acquired at a speed of 15 frames/min. It is clearly demonstrated in Fig. 3 (b) that for a silica microtoroid, although the resonance of each mode of the doublet is significantly affected by the change in temperature, the amount of splitting between the split modes is not affected, similar to what has been observed in polymer coated silica microtoroid. This implies that regardless of the material of microresonators, mode splitting is robust against thermal perturbations. Therefore, we conclude that perturbations in the environmental temperature have a negligible influence on the mode splitting based sensor schemes with respect to the sensing schemes based on detecting a single resonance.

In conclusion, we have demonstrated scatterer induced mode splitting in polymer-silica composite microresonators. Temperature response of the split resonances in polymer coated and silica microresonators is investigated to show the robustness of mode splitting to thermal fluctuations. Similarly, the self-referencing property of mode splitting suggests that the influence of interfering or undesired environmental perturbations (e.g., humidity and air pressure) which affects the whole resonant system uniformly can be avoided. As mode splitting is affected by the overlap between the particle and the WGM field, any changes in the overlap (e.g., acoustic pressure induced deformation of the polymer) and properties of the particle (e.g., refractive index or shape) will lead to variations in the splitting, which in turn, can be utilized to detect these changes. Demonstration of mode splitting in polymer coated resonators and the robustness of its spectrum against perturbations that affect the microscale resonator homogenously opens exciting possibilities in resonator-based sensing technologies by combining the versatility of polymers with the self-referenced and highly-sensitive mode splitting technique.

The authors acknowledge support from MAGEEP (McDonnell Academy Global Energy & Environment Partnership) and CMI (Center for Materials Innovation) at Washington University in St. Louis.